\documentclass[prl,twocolumn,amsmath,amssymb,amsfonts,superscriptaddress]{revtex4-2}

\usepackage{graphicx}
\usepackage{siunitx}
\usepackage{mathtools}

\usepackage{xcolor}
\usepackage{comment}

\usepackage{glossaries}
\newacronym{itt}{ITT}{integration through transients}
\newacronym{mct}{MCT}{mode-coupling theory of the glass transition}
\newacronym{fem}{FEM}{finite-element method}
\newacronym{lb}{LB}{Lattice-Boltzmann}
\newacronym{ce}{CE}{constitutive equation}
\newacronym{ice}{ICE}{integral constitutive equation}
\newacronym{am}{AM}{additive manufacturing}
\newacronym{mems}{MEMS}{micro-electromechanical systems}
\let\bs\boldsymbol
\DeclareMathOperator\tr{tr}
\usepackage{CJK}

\begin{document}
\def\dlr{\affiliation{Institut f\"ur Materialphysik im Weltraum,
  Deutsches Zentrum f\"ur Luft- und Raumfahrt (DLR) e.V., 51170 K\"oln, Germany}}
\def\hhu{\affiliation{Heinrich-Heine-Universit\"at D\"usseldorf,
  Universit\"atsstra{\ss}e~1, 40225 D\"usseldorf, Germany}}
\def\tud{\affiliation{Institut f\"ur Angewandte Mathematik (LS3),
  Technische Universit\"at Dortmund, Vogelpothsweg~87, 44227 Dortmund, Germany}}

\title{Residual stresses couple microscopic and macroscopic scales}

\author{Sebastian Steinh\"auser}\dlr\hhu
\author{Timm Treskatis}\tud
\author{Stefan Turek}\tud
\author{Thomas Voigtmann}\dlr\hhu

\date\today

\begin{abstract}
We show how residual stresses emerge in a visco-elastic material
as a signature of its past flow history, through an interplay between
flow-modified microscopic relaxation and macroscopic
features of the flow.
Long-lasting temporal-history dependence of the microscopic dynamics
and nonlinear rheology are incorporated through the \gls{mct}.
The theory's \gls{ice} is coupled to continuum mechanics in
a \gls{fem} scheme that tracks the flow history through the Finger tensor.
The method is suitable for a calculation of residual stresses from
a ``first-principles'' starting point following well-understood
approximations.
As an example, we calculate within a schematic version of \gls{mct}
the stress-induced optical birefringence
pattern of an amorphous solid cast into the shape of a slab with
a cylindrical obstacle and demonstrate how \gls{fem}-\gls{mct} can
predict the dependence of material properties on the material's
processing history.
\end{abstract}

\maketitle
\glsresetall

Materials that flow carry stresses, and once such a material solidifies,
some of those stresses can remain in the solid as flow-induced
residual stresses.
While the textbook elastic solid is assumed to be stressed only
if strained, the possibility of
internal stresses has been discussed theoretically
since at least the 1930's \cite{Reissner.1931}.
All continuum mechanics requires, is that the stress tensor $\bs\sigma$
inside the material is self-equilibrated, $\vec\nabla\cdot\bs\sigma=0$,
which in general allows non-zero stresses at rest.

Residual stresses are now recognized as a major factor controlling the
mechanical \cite{Withers.2007,Hirobe.2021} and optical \cite{Wissuchek.1999,Yablon.2004,Yi.2011,McMillen.2016} performance of materials and composites \cite{Parlevliet.2006,*Parlevliet.2007}. They arise from rapid
solidification, through chemical means, or from spatial variations in the
material coefficients.
Spectacular demonstrations of their effect
go back to exploding glass droplets
by Prince Rupert in the 1660's \cite{Aben.2016,Brodsley.1986,Chandrasekar.1994},
and perhaps \cite{em:Berry.2006} to the images produced by ``magic mirrors''
(\begin{CJK*}{UTF8}{gbsn}^^e9^^80^^8f^^e5^^85^^89^^e9^^95^^9c\end{CJK*})
of the Western Han dynasty around 200~B.C.\ \cite{rheology:Murray.1987,unsorted:Yan.1992}. The controlled use of
residual stresses allows to produce improved materials \cite{Zhang.2006},
such as scratch-resistant smartphone covers or safety glasses.
The stability of thin films and coatings, produced from a strongly
flowing state, depends on control of residual stresses \cite{Reiter.2005};
in \gls{mems} they have decisive influence on the device's function
\cite{Schloegl.2023}.
In turn, coatings allow to control residual stresses, e.g.,
in additively manufactured biocompatible implants \cite{Hein.2022}.
Residual stresses play an important factor in maintaining
homeostatic states in living tissues \cite{Fung.1990,Lanir.2009,Lanir.2017,Pena.2015,Ciarletta.2016,Sigaeva.2019}.
Finally, \gls{am} has re-emphasized
the role of residual stresses
in determining the mechanical properties and failure behavior of the
final piece \cite{Mercelis.2006,Patterson.2022,ChenW.2019}.
Despite their ubiquitous nature and importance, residual stresses remain
hard to measure \cite{Withers.2001a,Withers.2001b,Schajer.2013,Guo.2021},
which necessitates improved models.

\begin{figure}
\includegraphics[width=\linewidth]{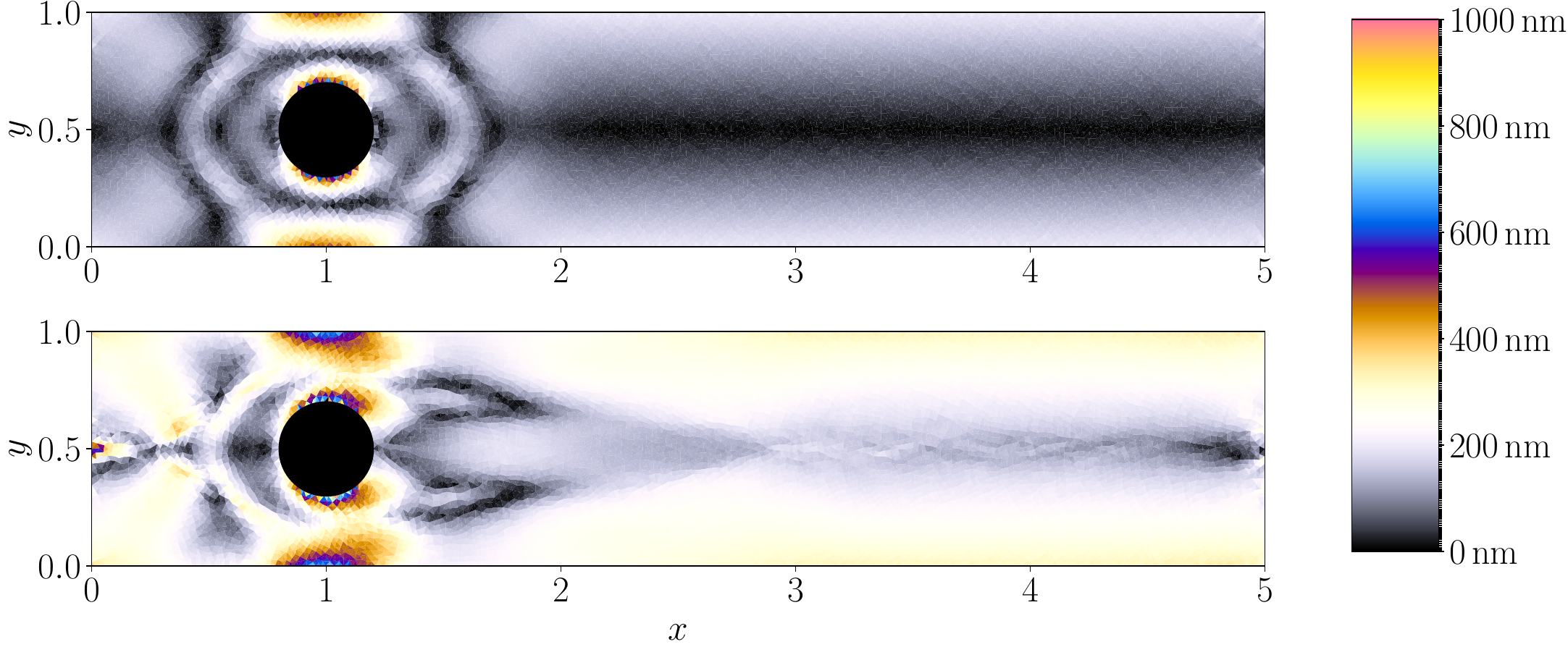}
\caption{\label{fig:biref}
  Residual-stress induced birefringence
  in a 5:1 rectangular (periodic) channel with a circular obstacle,
  after cessation of a pressure-driven flow with pressure drops
  (a) $\Delta p/G_0=1$, (b) $\Delta p/G_0=10$.
  Colors represent optical path lengths for white-light illumination
  (as indicated by the color bar)
  assuming sample thickness $h$ and stress-optical coefficient $C$
  such that $CG_0h=\SI{80}{\micro\meter}$ respectively
  $\SI{800}{\micro\meter}$,
  following a standard CIE color model (see text for details).
}
\end{figure}

We aim to explain residual stresses in amorphous materials,
in particular colloidal materials \cite{Ballauff.2013,Mohan.2013}
after cessation of flow, in a theoretical framework that is founded in
microscopic principles:
the combination of \gls{mct} with the \gls{fem}.
The FEM-MCT model predicts 
the appearance of spatial patterns of residual stresses
in an amorphous solids that depend on both the flow
history and geometry.
For example, the morphology of stresses remaining around a cylindrical
obstacle in a channel depends crucially on the pressure gradient that was
used to drive the flow before it came to rest (Fig.~\ref{fig:biref}).
It reveals an interesting interplay betwen
macroscopic symmetries of the flow, and microscopic relaxation
response, \textit{i.e.}, a true \emph{multi-scale} phenomenon:
whether and how many residual stresses remain, depends
on how the microscopic relaxation is modified by, and influences the,
macroscopic flow.

Predicting flow-induced residual stresses in glassy materials
from first principles is not a simple task:
they
remain because the structural relaxation
dynamics of the material becomes
extremely slow, and moreover the microscopic relaxation mechanisms are
strongly influenced by the macroscopic flow conditions (rendering it a
problem of nonlinear rheology).
The phenomenon is thus intrinsically a non-equilibrium, non-linear response
one:
it is the relaxation from a flowing
to the quiescent state that gets ``stuck'' in a stressed out-of-equilibrium
state.
This also implies that residual stresses depend on the past processing
history of a material. Thus, a description of the material in its final
state is incomplete, and one requires a model that includes the complete
flow history in its constitutive equation.
There is no separation of time scales between the microscopic relaxation
and the macroscopic flow.
This is a much more complicated
situation than for example the modeling of thermal residual stresses
\cite{Wei.2021,Chen.2008,Strantza.2018,Chen.2020,Grilli.2021}.

The accepted theory of macroscopic flow is formulated through the
Navier-Stokes equations for the mass density
$\rho(t,\vec x)$ and the flow velocity $\vec v(t,\vec x)$,
\begin{equation}\label{eq:ns}
  \rho\frac{D}{Dt}\vec v=-\vec\nabla p+\vec\nabla\cdot\bs\sigma
\end{equation}
where $D/Dt=\partial_t+\vec v\cdot\vec\nabla$ is the advected derivative
and $p$ is the pressure applied to the system.

The stress tensor $\bs\sigma$ encodes the dissipative terms that arise
in the coarse-graining implicit in Eq.~\eqref{eq:ns}. They need to be
described by a material law, in the form of a \gls{ce} that links the
stress tensor back to the fields, most notably to suitably frame-invariant
combinations of the velocity gradients $\bs\kappa=\vec\nabla\vec v$.
Most \gls{ce} used in the literature
are empirical, encoding the microscopic details of a material in a few
ad-hoc parameters.
Here, we will follow an approach that rests on a microscopic theory for
$\bs\sigma$ for glass-forming fluids. It will be useful to split
$\bs\sigma=\bs\sigma_N+\bs\sigma_P$, where $\bs\sigma_N=\eta_\infty(\bs\kappa
+\bs\kappa^T)$ describes a Newtonian background viscosity of the fluid.

For slow structural relaxation in glass-forming fluids, a
successful microscopic theory is the \gls{mct},
specifically its extension to (colloidal) rheology \cite{Fuchs.2002c,Brader.2007,Brader.2008,Fuchs.2009}.
Using the \gls{itt}, a general non-equilibrium statistical-physics
framework, one derives a material law of the form
\begin{equation}\label{eq:sigmapL}
  \bs\sigma_P(t)=\int_{-\infty}^t{\mathbb G}(t,t',[\bs B]):[-\partial_{t'}\bs B_{tt'}]\,dt'\,,
\end{equation}
where the flow enters through the Finger tensor
$\bs B_{tt'}$, the rotation-invariant measure
of deformations occurring between time $t'$ and $t\ge t'$.
Standard continuum-mechanics arguments imply \cite{Gurtin.2009}
\begin{equation}\label{eq:Btt}
\overset\triangledown{\bs B}_{tt'}=\frac{D}{Dt}{\bs B}_{tt'}-\bs\kappa_t\cdot
\bs B_{tt'}-\bs B_{tt'}\cdot\bs\kappa^T_t=\bs0\,.
\end{equation}

The generalized non-linear-response shear modulus $\mathbb G$
(a fourth-rank tensor) encodes the microscopic stress relaxation
and bears a functional dependence on $\bs B$ representing the flow history.
Equation~\eqref{eq:sigmapL} is the general form expected from
statistical physics, as it describes
the time-delayed response of a system to a past perturbation.
In rheological terms, it is an \gls{ice} that---in contrast to most
empirical laws---cannot be rewritten as a
differential equation (because of the functional dependence of $\mathbb G$
on the flow).
\Gls{ce} are not in general partial differential equations, because they are not
local conservation laws in contrast to the fundamental field-theory laws.
This poses a major complication for the numerical approach together with
the \gls{fem}, but it also encodes physically relevant flow-history
phenomena, such as those giving rise to residual stresses.

\Gls{itt} provides a microscopic expression for $\mathbb G$
upon which \gls{mct} constructs an approximate \gls{ce}
derived from first principles,
and predicts the appearance of residual stresses \cite{Ballauff.2013}.
It works under the assumption that the flow is slowly varying
in space and that stress relaxation is dominated by the relaxation of
density fluctuations on the length scale of interparticle separation.
We consider low-Reynolds-number flow on long time scales;
for this reason the formulation of \gls{mct} for colloidal suspensions is
sufficient.
In order to highlight the qualitative features predicted by \gls{mct},
we proceed with an established schematic model \cite{Brader.2009} that
approximates the shear modulus by an isotropic form,
$G_{ijkl}(t,t',[\bs B])=\delta_{ik}\delta_{jl}G(t,t',[\bs B])$.
Further approximations replacing $G$ by a simple decay function,
$G(t,t')\approx G_\infty\exp[-(t-t')/\tau]$, say, would recover from
Eq.~\eqref{eq:sigmapL} the upper-convected Maxwell model for $\bs\sigma_P$,
or equivalently, the Oldroyd~B model for $\bs\sigma$.
This type of model misses two ingredients to predict history-dependent
residual stresses, that \gls{mct} adds: $G$ is not
time-translational invariant, and it needs to depend on
the flow history itself.

Technically, schematic \gls{mct} encodes time-delayed
response in a non-Markovian model for the dominant density-relaxation
mode $\phi(t,t')$. Neglecting macroscopic gradients in this dynamics
\cite{Nicolas.2016}, $\vec v\cdot\vec\nabla\phi\approx0$, we set
$G(t,t',[\bs B],\vec x)\approx G_0\phi^2(t,t',[\bs B],\vec x)$ and
\cite{Brader.2009}
\begin{subequations}\label{eq:phim}
\begin{gather}
\begin{multlined}[t][.8\linewidth]
\label{eq:phi}
  \tau_0\partial_t\phi(t,t',\vec x)+\phi(t,t',\vec x)\\
  +\int_{t'}^tm(t,t'',t',\vec x)\partial_{t''}\phi(t'',t',\vec x)\,dt''=0\,,
\end{multlined}\\
\label{eq:m}
  m(t,t'',t')=h_{tt''}[\bs B]h_{tt'}[\bs B]\left(v_1\phi(t,t'')+v_2\phi^2(t,t'')\right)\,,
\end{gather}
dropping notation of $\vec x$- and $\bs B$-dependence as convenient.
The latter comes in through the memory-loss functions,
$h_{tt'}[\bs B]=1/(1+\tr(\bs B_{tt'}-\bs1)/\gamma_c^2)$ that describe decorrelation
of microscopic density fluctuations by flow advection.
The parameter $\gamma_c=1/10$ fixes a strain scale to connect to
microscopic units.
\end{subequations}

The core of \gls{mct} is the approximation Eq.~\eqref{eq:m}. In the
schematic model, $(v_1,v_2)$ play the role of the coupling coefficients
that within the full theory are given by the equilibrium static
structure functions of the fluid. At a certain coupling strength (\textit{e.g.},
driven by density or temperature), the solutions of Eq.~\eqref{eq:phim}
turn \emph{glassy}: there is a set of critical points $(v_1^c,v_2^c)$
separating fluid-like solutions characterized by
$\lim_{t\to\infty}\phi(t,t',[\bs B=0])=0$ from glass-like ones,
$\lim_{t\to\infty}\phi(t,t',[\bs B=0])=f>0$.
Setting $(v_1,v_2)=(v_1^c+\epsilon/(\sqrt2-1),v_2^c)$
with $v_2^c=2$ and $v_1^c=2(\sqrt2-1)$
ensures the asymptotic
features of the model to match that found by fully microscopic calculations
for typical colloidal suspensions \cite{Brader.2009}.

We fix $\epsilon=1/100$; this implies that we study a material
that is glassy at rest, but has a yield stress of $\sigma_y=G_\infty\gamma_c$
with $G_\infty=G_0f^2\approx G_0/10$, and will be fluidized locally when
the stress is above this value.
The model predicts shear thinning,
common empirical tensorial yield criteria \cite{Brader.2009},
and residual stresses upon instantaneous cessation of simple-shear
flow \cite{Ballauff.2013,Fritschi.2014}.

The results shown here have been obtained by a combined
numerical scheme that uses the finite-element method (FEM) augmented
with a fully nonstationary \gls{mct} solver (adapted from Ref.~\cite{Voigtmann.2012}).
A Marchuk-Yanenko splitting disentangles the calculation
of $(\vec v,p)$ [Eq.~\eqref{eq:ns}] from that of $(\bs B,G)$
[Eqs.~\eqref{eq:sigmapL}--\eqref{eq:phim}]
into separate iterative time steps.
All time-derivatives are
discretized with a simple implicit Euler scheme. For each $t'<t$
the Finger tensors are approximated as piecewise constant in space and determined from
Eq.~\eqref{eq:Btt}. The stresses are then evaluated following the solution of Eqs.~\eqref{eq:phim}.
Since the relaxation functions extend far back in time, but also become
slowly varying, a pseudo-logarithmic grid in $t-t'$ is employed.
The updated
stress contribution $\bs\sigma_P$ then enters Eq.~\eqref{eq:ns},
which is solved at each time step with a standard $P_2$--$P_0$
finite-element scheme, using second-order continuous Lagrange elements
(CG2) for the velocity, and discontinuous zeroth-order elements (DG0)
for stress and pressure, ensuring correct cellwise mass and momentum conservation.
The entire code is implemented in the programming language Python,
using the package FEniCS (version 2019.2) \cite{Logg.2012,FEniCS}
for the solution of the discretized Navier-Stokes equations, employing the
MUMPS LU solver.
In comparison to previous
attempts of integrating \gls{mct} with continuum mechanics based on the
\gls{lb} technique \cite{Papenkort.2014,*Papenkort.2015,*Papenkort.2015b},
\gls{fem} offers a more straight-forward approach to deal with non-Newtonian
flows in complex geometries and using non-uniform meshes.

We consider two-dimensional incompressible ($\vec\nabla\cdot\vec v=0$)
flow in an infinite rectangular channel (height $H$,
periodicity $L=5H$ along the $x$-direction) that is initially driven
by a pressure gradient,
$\vec\nabla p(t)=-(\Delta p/L)\vec e_{\vec x}$ for $0<t<t_\text{off}$.
Periodic boundary conditions are applied in the $x$-direction, no-slip
boundary conditions on the walls perpendicular to the $y$-direction.
The pressure gradient is removed,
$\vec\nabla p(t)=0$, for $t>t_\text{off}$.
To break the trivial $x$-translation symmetry of the problem,
a circular obstacle (radius $0.2H$, no-slip b.c.) is placed centered in the channel
at $(x_0,y_0)=(1,0.5)H$.

Setting $G_0=1$, $\eta_\infty=1$, and $\tau_0=1$ fixes the remaining units.
The time step for the numerical solution is $dt=0.05$, on a
grid with $128$ mesh cells across its diameter.
With these parameters, a single run (2032~timesteps) takes around
10~days utilizing 24~cores of a dual Xeon E5-2650v4 workstation
and around 45~GB of memory.

\begin{figure}
\includegraphics[width=\linewidth]{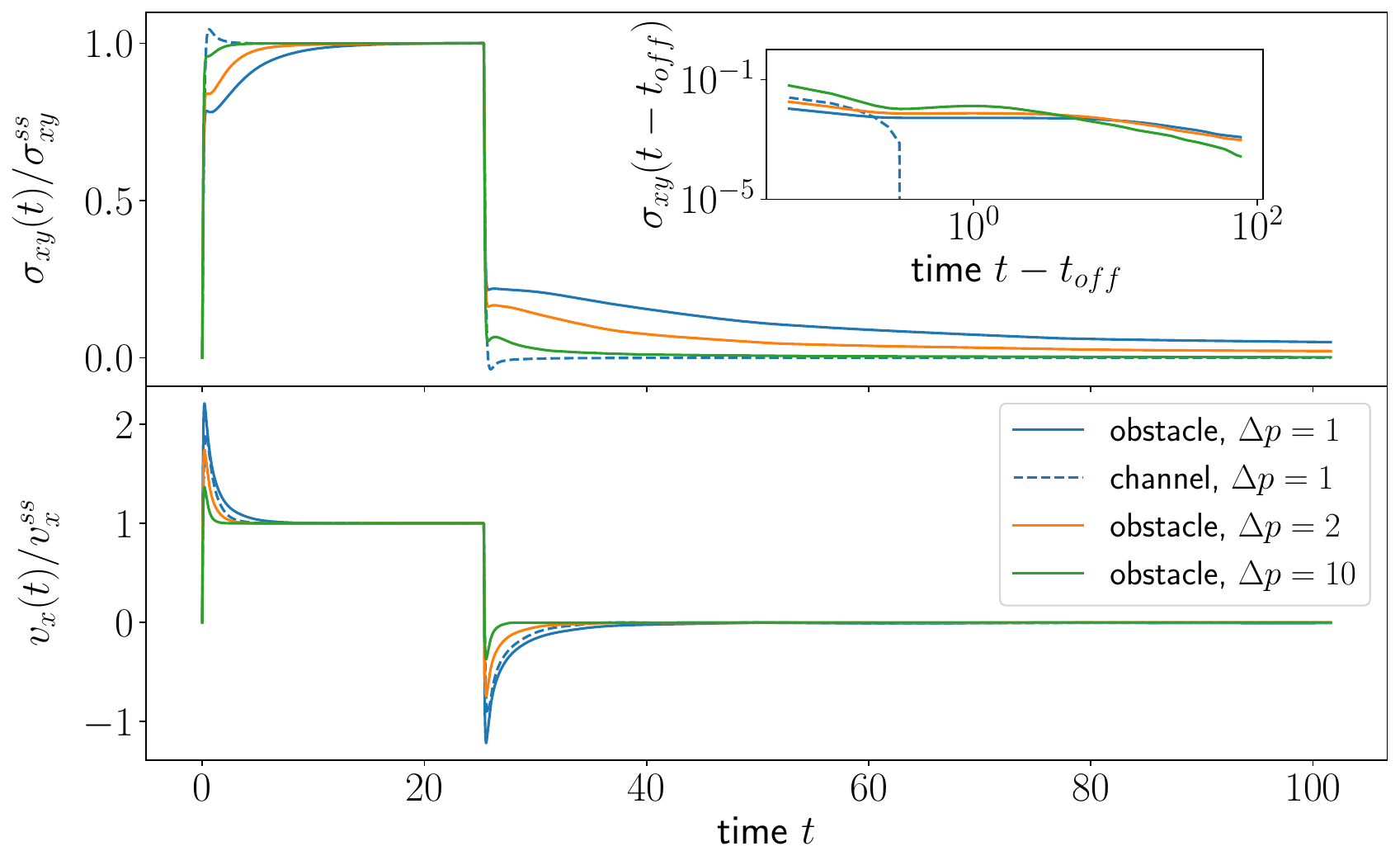}
\caption{\label{fig:stress_time}
  Shear stress $\sigma_{xy}(t)$ and velocity $v_x(t)$ as a function of time
  in the 5:1 rectangular channel with cylindrical obstacle (solid lines),
  for different initial driving pressures $\Delta p/G_0$ as indicated
  (active for $0<t<t_\text{off}=25.4\tau_0$).
  The inset provides a logarithmic zoom of the stress decay.
  Dashed lines correspond to a channel without obstacle.
}
\end{figure}

We prepare initial flowing states by applying a constant pressure gradient
to the stress-free material at rest;
the driving force is switched off instantaneously at $t_\text{off}=25.4\tau_0$.
The velocity initially increases towards the stationary state through
a distinctive overshoot typical for the startup flow of viscoelastic
fluids \cite{Papenkort.2015}; the stresses concomittantly increase towards
their stationary values on a somewhat longer time scale
(Fig.~\ref{fig:stress_time}).

After removal of the pressure gradient, the velocity displays a
pronounced undershoot before relaxing towards zero.
This can be rationalized as another
viscoelastic effect \cite{Papenkort.2015}: the presence of stresses
causes the fluid to be driven in the direction opposite of the intial
flow direction, until the stresses have sufficiently relaxed.

Crucially, we observe the shear stress $\sigma_{xy}(t)$ to completely
relax in the channel without obstacle (dashed line in Fig.~\ref{fig:stress_time}),
while in the presence of the obstacle, the velocity back-leash is
much less pronunced, and finite shear stresses remain even for times where
the velocity has already decayed to zero. These are geometry-dependent
residual stresses.
The difference arises from the coupling of the macroscopic deformation
gradients into the microscopic relaxation, Eqs.~\eqref{eq:phim}.
The macroscopic momentum balance at rest,
$\vec\nabla\cdot\bs\sigma=0$,
implies $\partial_y\sigma_{xy}=-\partial_x\sigma_{xx}$, and thus
any residual shear stress arising from a flow that is translational invariant
along the $x$-direction, needs to obey $\sigma_{xy}(y)=\text{const.}$
For homogeneous simple shear between parallel plates,
Eqs.~\eqref{eq:phim} in the glass
($\epsilon\ge0$) predict a
finite residual shear stress $\sigma_{xy}(t\to\infty)=\sigma_\infty>0$
\cite{Ballauff.2013,Fritschi.2014}, automatically compatible
with $\vec\nabla\cdot\bs\sigma=0$.
The $x$-translationally invariant channel flow imposes
$\sigma_{xy}(y=0)=0$ by symmetry, so that here $\sigma_\infty=0$ is the only
solution compatible with the macroscopic momentum balance.
If however the translational symmetry is broken (e.g., by the obstacle),
non-zero residual shear stresses become permissible again.
This is the non-trivial solution obtained from FEM-MCT.

To connect to typical measurements, Fig.~\ref{fig:biref} presents stress patterns as if
observed for an optically stress-induced birefringent medium between
crossed circular polarizers. This polarimeter setup is commonly used
as a non-destructive method to assess
internal stresses in optically transparent media \cite{Nelson.2021}.
It rests on the stress-optical effect: otherwise non-birefringent
materials become optically birefringent in response to
mechanical stress. To first order, there holds Maxwell's stress-optical law
\cite{Maxwell.1853}, $\Delta n=n_o-n_e=C(\sigma_1-\sigma_2)$, where $\sigma_{1,2}$
are the stress eigenvalues in the plane perpendicular to the propagation
of light, and $n_{o,e}$ are the refractive indices along the optical
axes defined by the corresponding stress eigenvectors. $C$ is called
the stress-optical coefficient.
Assuming a slab of material whose properties are invariant along the
light-propagation direction (taken to be along $z$), a textbook calculation
shows that the transmitted light intensity when placing the sample between
two circular polarizers, is $I(\lambda,\delta)=I_0(\lambda)\cos(\delta/2)^2$, where
$\delta=(2\pi z/\lambda)\Delta n$ is the (stress-dependent) optical retardation.
If we assume illumination by a white-light source ($I_0$ constant across
the visible spectrum $\lambda\in[360,830]\text{nm})$, a
colorful transmission spectrum will be recorded by the observer. This
spectrum $I(\lambda,\delta)$ can be converted to empirical RGB values
of colors as perceived by the human eye \cite{Sorensen.2013}. This is
the color bar used in Fig.~\ref{fig:biref}.


\begin{figure}
\includegraphics[width=\linewidth]{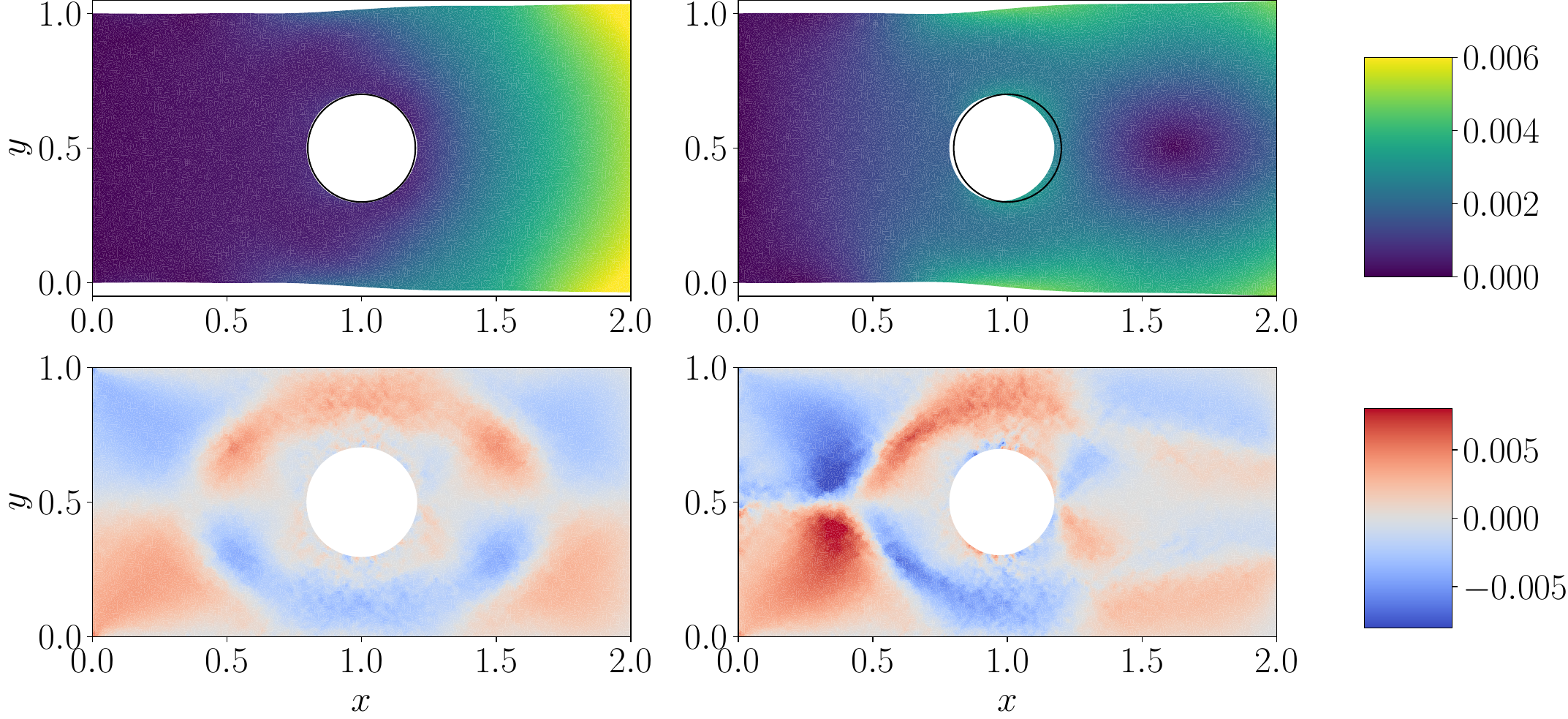}
\caption{\label{fig:eigenstrain}
  Top: Eigenstrain balancing the residual stresses
  after cutting a section of width $\Delta x=2$ from the channel, for
  initial pressure drop $\Delta p/G_0=1$ (left) and $\Delta p/G_0=10$
  (right). The magnitude is shown as color, the deformation of the
  sample is shown amplified by a factor $10$ respectively $100$.
  Bottom: corresponding relaxed residual shear stress $\hat\sigma_{xy}$
  (result for $\Delta p/G_0=10$ left magnified by a factor $10$).
}
\end{figure}

To illustrate the effect of the residual stresses on the mechanical properties of
the final material, let us consider the glass as a linear-elastic material
with stress tensor $\bs\sigma=\bs\sigma_\text{res}+\hat{\bs\sigma}(\bs\varepsilon)$,
composed of a residual stress and a strain-dependent elastic stress.
The deformation field $\vec u$ in the material's bulk
volume $\Omega$ solves
\begin{subequations}\label{eq:linel}
\begin{gather}
  \vec\nabla\cdot\left[\hat{\bs\sigma}+\bs\sigma_\text{res}\right]=0
  \qquad\text{in $\Omega$}\,,\\
  \hat{\bs\sigma}=\lambda(\tr\bs\varepsilon)\bs1+2\mu\bs\varepsilon\,,\\
  \bs\varepsilon=(\vec\nabla\vec u+(\vec\nabla\vec u)^T)/2\,.
\end{gather}
\end{subequations}
Here, $\bs\sigma_\text{res}=\bs\sigma_P(t\to\infty)$ is approximated by
the final-time stress tensor obtained from FEM-MCT. The set of Eqs.~\eqref{eq:linel} is solved by a standard \gls{fem} simulation with
$\mu=G_0$ and $\lambda=1.5G_0$, using CG1 elements for the displacement field.
A truly
self-equilibrated stress, $\vec\nabla\cdot\bs\sigma_\text{res}=0$ does
not enter linear elasticity at all. However, this hinges on the
boundary conditions: for $\bs\sigma_\text{res}$ generated by the channel flow
inside fixed walls, these correspond to $\vec u=0$ on $\partial\Omega$.
For the sake of demonstration, let us ``cut'' the solid from its casting
shape, i.e., change to free boundary
conditions, $\bs\sigma\cdot\vec n=0$ on $\partial\Omega$
except on vertical walls at $x\in\{0,2H\}$ (where $\vec u=0$ is imposed).
It implies
$\hat{\bs\sigma}\cdot\vec n=-\bs\sigma_\text{res}\cdot\vec n$ and an
ensuing non-trivial deformation field $\vec u$ in $\Omega$.
These eigenstrains caused by the residual stresses are shown in
Fig.~\ref{fig:eigenstrain}(a,b) for the two different residual-stress
levels; the comparison brings out that they are indeed noticably different,
\textit{i.e.}, the material has different elastic properties due to its
different past-flow history. The eigenstrains relax some of
these stresses, but still, relaxed residual stresses remain, as shown
in Fig.~\ref{fig:eigenstrain}(c,d).

Both the unrelaxed (Fig.~\ref{fig:biref}) and relaxed (Fig.~\ref{fig:eigenstrain})
stresses show an interesting viscoelastic symmetry-breaking effect:
Since the stationary Stokes-flow solution for a Newtonian fluid is
time-reversal symmetric (the basis of the famous no-swimming theorem
at low Reynolds number \cite{Purcell.1977}), the corresponding
flow patterns obey an inflection symmetry around the obstacle.
Our results for small $\Delta p/G_0$ demonstrate this.
But for large $\Delta p/G_0$, viscoelastic terms from the \gls{ice}
break this symmetry, even in zero-Reynolds-number flow.
It appears that the residual stresses exhibit the pattern of
``viscoelastic turbulence'' \cite{Groisman.2000,Larson.2000,Morozov.2007}
even though typically shear thinning is observed to suppress the
effect in stationary flow \cite{Casanellas.2016,Nicolas.2016}.

To conclude, we presented predictions of residual-stress patterns emerging
in non-trivial flow geometry, based on a microscopic theory of the flow
of glass-forming fluids, the \gls{mct}, combined with \gls{fem} simulations
that allow to bridge from the microscopic to the macroscopic scale.
We see this as a first conceptual step towards a scale-bridging approach
modeling a material's property through its entire processing history.

The cessation of pressure-driven channel flows reveals an intriguing
coupling of microscopic and macroscopic scales:
While \gls{mct} in principle allows for residual stresses simply due to the
infinite microscopic structural relaxation time in the ideal glass,
continuum mechanics enforces
those stresses to be macroscopically self-equilibrated, i.e.,
$\vec\nabla\cdot\bs\sigma=0$ at rest.
Through this, the macroscopic symmetries and boundary conditions
enter the microscopic relaxation dynamics and change it qualitatively.

\begin{acknowledgments}
This work has been funded by the Deutsche Forschungsgemeinschaft
(DFG, German Research Foundation) -- project number 431117597.
\end{acknowledgments}

\bibliographystyle{apsrev4-2}
\bibliography{references}
\end{document}